\documentclass[letterpaper,aps,prd,superscriptaddress,showpacs,nofootinbib,floatfix,twocolumn]{revtex4}
\usepackage{graphicx,epsf,epsfig,color,bm}
\def\Journal#1#2#3#4{{#1}{\bf #2}, #3 (#4)}

\def\JPG{{J. Phys}~{\bf G}}

\def\NIMA{{Nucl. Instr. Meth.}~{\bf A}}

\def\PRL{Phys. Rev. Lett.\ }

\def\PRC{{Phys. Rev.}~{\bf C}}

\newcommand{\sqrts}{\mbox{$\sqrt{s}$}}

\newcommand{\dA}{\mbox{\textit{d}+A}}
\renewcommand{\AA}{\mbox{A+A}}

\newcommand{\pp}{\mbox{\textit{p}+\textit{p}}}

\newcommand{\pT}{\mbox{$p_T$}}

\newcommand{\gevc}{\mbox{${\mathrm{GeV/}}c$}}

\newcommand{\jpsi}{\mbox{$J/\psi$}}
\newcommand{\epem}{\mbox{$e^+e^-$}}

\begin{document}


\title{High $p_{T}$ non-photonic electron production in $p$+$p$ collisions at $\sqrt{s}$ = 200 GeV}
\input{makeLatex.php}
\date{\today}


\begin{abstract}
    We present the measurement of non-photonic electron production at high transverse momentum ($p_T > $ 2.5 GeV/$c$) in $p$+$p$ collisions at $\sqrt{s}$ = 200 GeV using data recorded during 2005 and 2008 by the STAR experiment at the Relativistic Heavy Ion Collider (RHIC).  The measured cross-sections from the two runs are consistent with each other despite a large difference in photonic background levels due to different detector configurations. We compare the measured non-photonic electron cross-sections with previously published RHIC data and pQCD calculations.  Using the relative contributions of B and D mesons to non-photonic electrons, we determine the integrated cross sections of electrons  ($\frac{e^++e^-}{2}$) at 3 GeV/$c < p_T <~$10 GeV/$c$ from bottom and charm meson decays to be ${d\sigma_{(B\to e)+(B\to D \to e)} \over dy_e}|_{y_e=0}$ = 4.0$\pm0.5$({\rm stat.})$\pm1.1$({\rm syst.}) nb and ${d\sigma_{D\to e} \over dy_e}|_{y_e=0}$ = 6.2$\pm0.7$({\rm stat.})$\pm1.5$({\rm syst.}) nb, respectively.

\end{abstract}

\pacs{13.20.Fc, 13.20.He, 25.75.Cj}

\maketitle

\section{Introduction}
\label{Introduction}
Heavy quark production in high-energy hadronic collisions has been a
focus of interest for years. It is one of the 
few instances in which experimental measurements can be compared with
QCD predictions over nearly the entire kinematical range
~\cite{Cacciari:2007hn,Cacciari:2005rk,Frixione:2005yf}. 
Due to the large masses of charm and bottom quarks, 
they are produced almost exclusively during
the initial high-Q parton-parton interactions and thus can be described by
perturbative QCD calculations.

Measurement of heavy flavor production in elementary
collisions represents a crucial test of the validity of the current
theoretical framework and its phenomenological inputs. 
It is also mandatory as a baseline for the interpretation of
heavy flavor production in nucleus-nucleus collisions
~\cite{Frawley:2008kk}.  In these heavy-ion collisions one investigates
the properties of the quark-gluon plasma (QGP), which is created at
sufficiently high center-of-mass energies. Many effects on
heavy flavor production in heavy-ion collisions have been observed but
are quantitatively not yet fully understood ~\cite{Frawley:2008kk}. Of
particular interest are effects which modify the transverse momentum
spectra of heavy flavor hadrons, including energy loss in the QGP ("jet
quenching")
~\cite{Mustafa:1997pm,Djordjevic:2005db,Wicks:2005gt,Djordjevic:2008iz,Sharma:2009hn},
as well as collective effects such as elliptic flow
~\cite{Moore:2004tg,vanHees:2005wb}.  In addition, $J/\psi$
might be regenerated in a dense plasma from the initial open charm yield 
~\cite{Grandchamp:2002iy}, making precise measurements
of the transverse momentum spectra in elementary $p$+$p$ collisions
imperative.

Open heavy-flavor production in \pp, \dA, and \AA\ collisions at
$\sqrt{s_{NN}}$ = 200 GeV has been studied at the Relativistic Heavy Ion
Collider (RHIC) using a variety of final-state observables
~\cite{Frawley:2008kk}. The STAR collaboration measured charm mesons
directly through their hadronic decay channels ~\cite{Dstar:2009df,
    Adams:2004fc,Baumgart:2008zz}. Due to the lack of precise
secondary vertex tracking and trigger capabilities these measurements
are restricted to low momenta ($\pT < 3$ GeV/$c$).  Both STAR
~\cite{Adams:2004fc, Abelev:2006db} and PHENIX
~\cite{Adler:2005xv,Adare:2006hc} also measured heavy flavor production
through semileptonic decays of charm and bottom mesons ($D, B
\rightarrow \ell\ \nu_{\ell}\ X$). While the measured decay leptons provide only limited information on the original kinematics of the heavy flavor parton, these measurements are facilitated by fast online triggers and extend the kinematic range to high \pT. 

In this paper, we report STAR results on non-photonic electron
production at midrapidity in \pp\  collisions at \sqrts~= 200 GeV using data recorded during year 2005 (Run2005) and year 2008 (Run2008) with a total integrated luminosity of
2.8 pb$^{-1}$ and 2.6 pb$^{-1}$, respectively. 
The present results are consistent with the Next-to-Leading Logarithm (FONLL) calculation within its theoretical uncertainties.
Utilizing the measured relative contributions of B and D mesons to non-photonic electrons which were obtained from a study of electron-hadron correlations (e-h)~\cite{e_h}, we determine the invariant cross section of electrons from bottom and charm meson decays separately at $p_T > $ 3.0 GeV/$c$. 

The article is organized as follows. In Sec.~\ref{Experiment} we describe the STAR
detectors and triggers relevant to this analysis. Sec.~\ref{Analysis} describes
the data analysis in detail, and in Sec.~\ref{result} we present and discuss the
results.  Sec.~\ref{conclusion} provides conclusions.

\section{Experiment}
\label{Experiment}
\subsection{Detectors}
\label{Detectors and Reconstruction}

STAR is a large acceptance, multi-purpose experiment composed of several
individual detector subsystems with tracking inside a large solenoidal
magnet generating a uniform field of 0.5 T~\cite{Ackermann:2002ad}. The detector subsystems
relevant for the present analysis are briefly described in the
following.
\subsubsection{Time Projection Chamber}
The Time Projection Chamber (TPC)~\cite{tpc} is the main charged particle tracking device
in STAR.  The TPC covers $\pm$1.0 units in pseudorapidity ($\eta$) for tracks crossing all layers of pads, and the full azimuth. Particle momentum is determined from track curvature in the solenoidal field.
In this analysis, TPC tracks are used for momentum determination, electron-hadron separation (using specific ionization $dE/dx$), to reconstruct the interaction vertex, and to project to the calorimeter for further hadron rejection. 
\subsubsection{Barrel Electromagnetic Calorimeter and \\ Barrel Shower Maximum Detector}

The Barrel Electromagnetic Calorimeter (BEMC) measures the
energy deposited by photons and electrons and provides a trigger
signal. It is located inside the magnet coil outside the TPC,
covering $|\eta| < 1.0$ and $2\pi$ in azimuth, matching the TPC
acceptance.  The BEMC is a lead-scintillator sampling electromagnetic
calorimeter with a nominal energy resolution of $\delta E/E \sim
14\%/\sqrt{E/1{\rm GeV}} \oplus 1.5\%$ ~\cite{bemc}. The full calorimeter is segmented into
4800 projective towers. A tower covers 0.05 rad in $\phi$ and 0.05
units in $\eta$.  Each tower consists of a stack of 20 layers of lead
and 21 layers of scintillator with an active depth of 23.5 cm.  The
first two scintillator layers are read out separately providing the
calorimeter preshower signal, which is not used in this analysis.  A
Shower Maximum Detector (BSMD) is positioned behind the fifth
scintillator layer.  The BSMD is a double layer wire proportional counter with strip readout.
The two layers of the BSMD, each containing 18000 strips, provide precise spacial resolution in $\phi$ and $\eta$ and improve the electron-hadron separation.
The BEMC also provides a high-energy trigger
based on the highest energy measured by a single tower in order to enrich the
event samples with high-$p_T$ electromagnetic energy deposition.

\subsubsection{Trigger Detectors}

The Beam-Beam Counters (BBC) ~\cite{bbc} are two identical counters located on each side of the interaction region covering the full azimuth and 2.1 $< |\eta| < $5.0. Each detector consists of sets of  small and large hexagonal scintillator tiles grouped into a ring and mounted around the beam pipe at a distance of 3.7 m from the interaction point. In both Run2008 and Run2005, the BBC served as a minimum-bias trigger to record the integrated luminosity by requiring a coincidence of signals in at least one of the small tiles (3.3 $< |\eta| < $5.0) on each side of the interaction region. The cross-section sampled with the BBC trigger is $26.1\pm0.2 ({\rm stat.})\pm1.8 ({\rm syst.})$ mb~\cite{bbc_cross_section} for $p$+$p$ collisions.  The timing signal recorded by the two BBC counters can be used to reconstruct the collision vertex along the beam direction with an accuracy of $\sim 40$ cm. 

The data in {\it d}+Au collisions recorded during year 2008 is used as a crosscheck in this analysis (see Sec~\ref{trg_eff}).
During this run, a pair of Vertex Position Detectors (VPD)
~\cite{vpd} was also used to select events. 
Each VPD consists of 19 lead converters plus plastic
scintillators with photomultiplier-tube readout that are positioned
very close to the beam pipe on each side of STAR. Each VPD
is approximately 5.7 m from the interaction point and
covers the pseudorapidity interval 4.24 $ < |\eta| <$ 5.1. 
The VPD trigger condition is similar to that of the BBC trigger except that the VPD has much better timing resolution, enabling the selected events to be constrained to a smaller range ($\sim\pm$30 cm in {\it d}+Au run) around the interaction point.

\subsection{Material Thickness in front of the TPC}
\label{material_budget}
Table~\ref{tab:material_run08} shows a rough estimate of material thickness between the interaction point and the inner field cage (IFC) of the TPC during Run2008 in the region relevant to the analysis. The amount of material is mostly from the beam pipe (BP), the IFC, air, and a wrap around the beam pipe. 
In Run2005 , the amount of material is estimated to be $\sim$ 10 times larger in front of the TPC inner field cage
 ~\cite{fujin_SQM08} and is dominated by the silicon detectors which were removed before Run2008. The contribution from the TPC gas is not significant because we require the radial location of the first TPC point of reconstructed tracks to be less than 70 cm (see Sec.~\ref{eID_cut_eff_est}) in the Run2008 analysis; furthermore, conversion electrons originating from TPC gas have low probability to be reconstructed by the TPC tracking due to the short track length.  While the Run2008 simulation describes the material distribution very well,  the material budget for the support structure and electronics related to the silicon detectors is not reliably described in the Run2005 simulation~\cite{bemc_cvs_cluster}.  This, however, has little effect on this analysis, as explained in Sec.~\ref{pho_reco_eff}.     

\begin{table} [htbp]
    \caption{\label{tab:material_run08}
        Estimates of material thickness of the beam pipe, the wrap around the beam pipe, the TPC inner field cage and air between the beam pipe and the inner field cage in Run2008.}
    \begin{ruledtabular}  \begin{tabular}{cccc}
           & source 			 & thickness in radiation lengths  			 \\
           \hline 
           & beam pipe              &  0.29~\% \\
          
           & beam pipe wrap       & $\sim$ 0.14~\%  \\
          
           & air       & $\sim$ 0.17~\% \\
          
           & inner field cage       & $\sim$ 0.45~\% \\
        \end{tabular} \end{ruledtabular}
\end{table}

\subsection{Triggers and Datasets}
\label{Triggers}

The data reported in this paper were recorded during Run2005 and Run2008
at \sqrts = 200 GeV.  All events used in this analysis
are required to satisfy a  BEMC trigger and a BBC
minimum-bias trigger.  In addition, event samples using a VPD trigger in the 2008 {\it d}+Au collisions are used for
systematic cross-checks as described in Sec.~\ref{trg_eff}.
 
To enrich the data sample with high-$p_T$ electromagnetic energy deposition, the BEMC trigger requires the energy deposition in at least one tower to exceed a preset threshold (high-tower). Most of
the energy from an electron or a photon will be deposited into a single
tower since the tower size exceeds the radius of a typical
electromagnetic shower. The Run2008 datasets used here were recorded using three 
high-tower triggers with different thresholds, corresponding to a sampled luminosity of $\sim$2.6 pb$^{-1}$.  Expressed in terms of transverse energy ($E_T$), the thresholds were approximately 2.6 GeV, 3.6 GeV
 and 4.3 GeV.  The Run2005 datasets used here are from two high-tower
triggers with $E_T$ thresholds of 2.6 GeV (HT1) and 3.5 GeV (HT2), corresponding to a sampled luminosity of $\sim$2.8 pb$^{-1}$. In the  Run2008 analysis,  datasets from different high-tower triggers are treated together after being combined, with double counting avoided by removing duplicates in the corresponding high-tower ADC spectra. Trigger efficiencies and prescale factors imposed by the data acquisition
system are taken into account during the combination. In the analysis of the Run2005 data, HT1 and HT2 data are treated separately. 

In Run2005 the integrated luminosity was monitored using the BBC minimum-bias trigger, while in Run2008, because of the large beam related background due to high luminosity, a high threshold high-tower trigger seeing a total cross section of 1.49 
$\mu $b, was used as luminosity monitor.

\section{Analysis}
\label{Analysis}
\subsection{Photonic Background Removal}
\label{Analysis_Principle}

The main background in this analysis 
is the substantial flux of photonic electrons from photon conversion in 
the detector material and Dalitz decay of $\pi^0$ and $\eta$ mesons. These contributions need to be subtracted 
in order to extract the non-photonic electron yield, which is 
dominated by electrons from semileptonic decays of heavy flavor mesons.

There are two distinct methods for evaluating contributions from photonic electrons. 
In the cocktail method, the estimated or measured invariant cross-sections are used to calculate contributions from various sources 
(mostly $\pi^0, \eta$ mesons),  
and to derive from those the photonic electron distributions. Given sufficient knowledge of the production yield of those mesons, this method allows one to determine directly the contributions from Dalitz decays. With this method, a detailed understanding of the material
distribution in the detector is required in order to evaluate the contribution from photon conversion. 
Another method, used in this analysis, is less dependent on the exact knowledge of the amount of material.  This method reconstructs the photonic electrons through the specific
feature that photonic electron-positron pairs have very small invariant mass. Not all photonic electrons can be identified this way since one of the electrons may fall outside of the detector acceptance, or has a very low momentum, in which cases both electrons in the pair are not reconstructed. This inefficiency must be estimated through simulation.

\begin{figure}[t]
    \centering
    \includegraphics[width=1.0\linewidth]{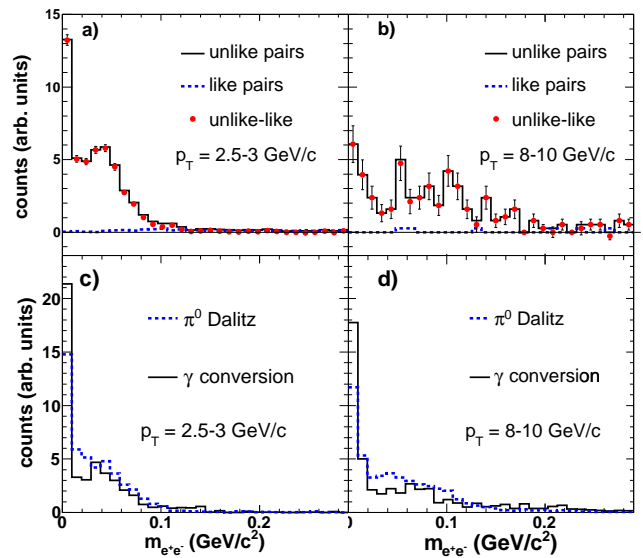}
    \caption{(Color online) The upper two panels show the electron pair invariant mass distributions for electrons at 2.5 GeV/$c < p_{T} <$ 3.0 GeV/$c$ (a) and at 8.0 GeV/$c < p_{T} <$ 10.0 GeV/$c$ (b). Solid and  dashed lines represent unlike-sign and like-sign pairs,  respectively. Closed circles represent the difference of unlike and like. The lower two panels show the simulated invariant mass spectra with electrons at 2.5 GeV/$c < p_{T} <$ 3.0 GeV/$c$ (c) and  at 8.0 GeV/$c < p_{T} <$ 10.0 GeV/$c$ (d). Solid and dashed lines represent results from $\gamma$ conversions and $\pi^0$ Dalitz decay.}
\label{fig:inv_mass}
\end{figure}

Electron pairs are formed by combining an electron with $p_{T} > 2.5$ GeV/$c$, which we refer to as a primary electron, with all other electrons (partners) reconstructed in the same event, with opposite charge sign (unlike-sign) or same charge sign (like-sign). The upper two panels of Fig.~\ref{fig:inv_mass} show the invariant mass spectra for primary electrons with 2.5 GeV/$c < p_{T} <$ 3.0 GeV/$c$ (left) and 8.0 GeV/$c < p_{T} <$ 10.0 GeV/$c$. The unlike-sign spectrum
includes pairs originating from photon conversion and Dalitz decay, as well as combinatorial background. The latter can be estimated using the like-sign pair spectrum. The photonic electron spectrum is obtained by subtracting like-sign from
unlike-sign spectrum (unlike-minus-like). The broad shoulders extending toward higher masses in the spectra are caused by 
finite tracking resolution, which leads to a larger reconstructed
opening angle when the reconstructed track helices of two conversion
electrons intersect each other in the transverse plane. The
overall width of the mass spectra depends on the primary electron $p_{T}$, 
but most photonic pairs are contained in range of $m_{ee} < 0.24$ GeV/$c^2$. 
The lower two panels of Fig.~\ref{fig:inv_mass} show the simulated invariant mass spectra of the two dominant sources of photonic electrons, $\pi^0$ Dalitz and $\gamma$ conversions, in the same two $p_T$ regions, which are similar in shape due to similar decay kinematics. 
The electron spectrum obtained from the unlike-minus-like method is from pure photonic electrons because the combinatorial background is accurately described by the like-sign pair spectrum according to the simulation and the simulated mass spectra are in qualitative agreement with the data.  This is also proved by the fact that the distribution of the normalized ionization energy loss (see Sec.~\ref{eID_cut_eff_est}) can be well described by a Gaussian function expected from electrons as shown in Fig.~\ref{fig:nsig_pure_e}.

We calculate the yield of non-photonic electrons according to $$ N(npe) =
N(inc) \cdotp \epsilon_{purity} - N(pho)/\epsilon_{pho}, $$ where
$N(npe)$ is the non-photonic electron yield, $N(inc)$ is the inclusive electron
yield, $N(pho)$ is the photonic electron yield, $\epsilon_{pho}$ is the
photonic electron reconstruction efficiency defined as the fraction of the
photonic electrons identified through invariant mass reconstruction, and
$\epsilon_{purity}$ is the purity reflecting hadron contamination in the
inclusive electron sample.

Electrons from open heavy flavor decays dominate non-photonic electrons.
The contribution from semi-leptonic decay of kaons is negligible for $p_T > $ 2.5 GeV/$c$~\cite{Adare:2006hc}. Electrons from vector mesons ($\rho, \omega, \phi$, $\jpsi$, $\Upsilon$) decays and Drell-Yan processes are subtracted from the measurement (see Sec.~\ref{feeddown} for details).

\subsection{Electron Reconstruction and Identification Efficiency }
\label{eID_cut_eff_est}
In the analysis of the Run2008 data, we select only tracks with $p_T > 2.5$ GeV/$c$ and $|\eta| < 0.5$. The event vertex position along the beam-axis ($V_z$) is required to be close to the center of the TPC, i.e. $|V_z|
< 30$~cm. To avoid track reconstruction artifacts, such as track splitting, the selected tracks are required to have at least $52\% $ of the maximum number of TPC points allowed in the TPC, a minimum of 20 TPC points and a distance-of-closest-approach (DCA) to the collision vertex less than 1.5 cm. For hadron rejection we apply a cut of $n\sigma_e > -1$ on the normalized ionization energy loss in the TPC~\cite{dedx}, which is defined as $$n\sigma_e = \log((dE/dx)/B_e)/\sigma_e, $$ where $B_e$ is the expected mean $dE/dx$ of an electron calculated from the Bichsel function~\cite{bichsel} and $\sigma_e$ is the TPC resolution of $\log((dE/dx)/B_e)$.

\begin{figure}[t]
    \centering
    \includegraphics[width=1.0\linewidth]{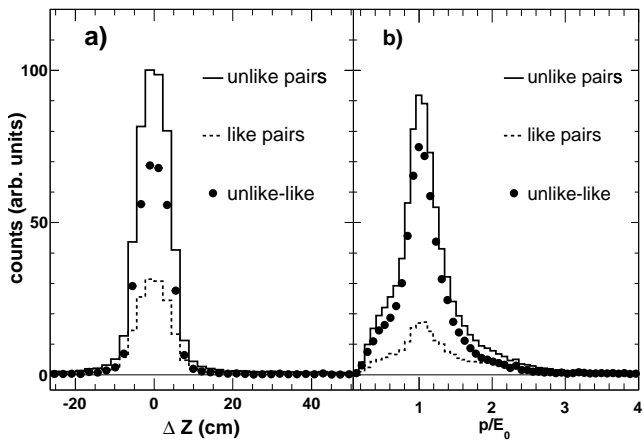}
    \caption{(a) The distribution of the minimum distance between an
        electron track projection point at the BEMC and all BEMC clusters
        along the beam direction from unlike-sign electron candidate pairs (solid line),
        like-sign electron candidate pairs (dashed line) and unlike-minus-like (closed circles).
        (b) $p/E_0$ distribution from unlike-sign electron candidate  pairs (solid line),
        like-sign electron candidate pairs (dashed line) and unlike-minus-like (closed circles).}
    \label{fig:bemc_assoc}
\end{figure}

We reconstruct clusters in the BEMC and the BSMD by grouping adjacent hits that are likely to have originated from the same incident particle~\cite{bemc_cvs_cluster}. The selected tracks are extrapolated to the BEMC and the BSMD where they are associated with the closest clusters. 
The association windows for electrons are determined by measuring the 
distance between
 the track projection point at the BEMC (BSMD) and the closest BEMC (BSMD) cluster using photonic electrons from the unlike-minus-like pairs. Figure~\ref{fig:bemc_assoc} (a) shows the distribution of this distance at the BEMC along the beam direction for electrons from unlike-sign, like-sign and unlike-minus-like pairs requiring $m_{e^{+}e^-} < 0.24 $ GeV/$c^2$,  a maximum 1.0 cm DCA
between two helical-shaped electron tracks and a 3.0 keV/cm $ < dE/dx <$ 5.0 keV/cm cut on ionization energy loss for partner tracks. Most electrons are inside a window of $\pm20$~cm around the track projection point. The
window in the azimuthal plane is determined to be $\pm0.2$ radian. Figure~\ref{fig:bemc_assoc} (b) shows the distribution of $p/E_0$ for electrons from unlike-sign, like-sign and unlike-minus-like pairs, where $E_0$ is the energy of the most energetic tower in a BEMC cluster. The distribution is peaked around one due to the small mass of the electron and the fact that most electron energy is deposited into one tower. We apply a cut of $p/E_0 < $~2.0 to further reduce hadron contamination. Cuts on the association with BSMD clusters are kept loose to maintain high efficiency. Each track is required to have more than one associated BSMD strip in both $\phi$ and $\eta$ planes.

\begin{figure}[b]
    \centering
    \includegraphics[width=1.0\linewidth]{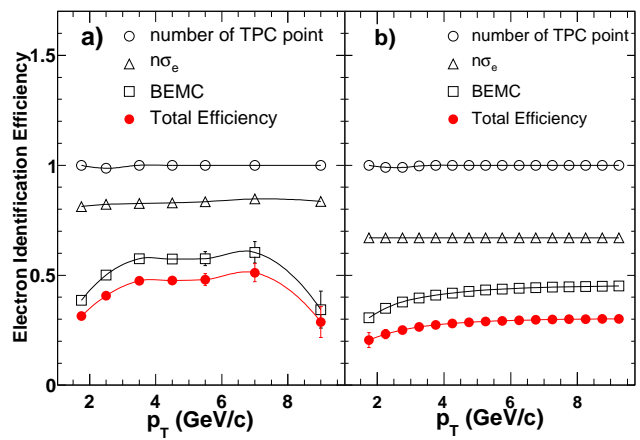}
    \caption{(Color online) Efficiencies of the cuts on number of TPC
        points (open circles), $n\sigma_e$ (open triangles) and BEMC
        (open squares) in (a) the Run2008 and (b) the Run2005 analyses.  The
        closed circles represent the total efficiency which is the
        product of all individual ones.}
    \label{fig:eID_eff}
\end{figure}

The efficiencies of electron identification
cuts are estimated directly from the data using pure photonic electrons obtained
from the unlike-minus-like pairs requiring $m_{e^{+}e^-} < 0.24 $ GeV/$c^2$,  a maximum 1.0 cm DCA
between two helical-shaped electron tracks and a 3.0 keV/cm $ < dE/dx <$ 5.0 keV/cm cut on ionization energy loss for partner tracks. A cut of $p_T > 0.3$ GeV/$c$ for partners is also applied, selecting a region where the simulation does a good job of describing the data. The efficiency for one specific cut is then
calculated as the ratio of electron yield before the cut to that
after the cut, while all the other electron identification cuts are
applied. To avoid possible correlation among different cuts,
efficiencies for all BEMC and BSMD cuts are calculated together.
Figure~\ref{fig:eID_eff} (a) shows the breakdown of the electron
identification efficiency as a function of $p_T$.  The drop in the low
$p_T$ region comes mainly from BSMD inefficiency, while the drop in the high $p_T$ region is caused by the $p/E_0$ cut. The uncertainties in the figure are purely statistical and are included as part
of the systematic uncertainties for the cross section calculation.

To maintain high track quality and suppress photonic electrons from conversion in the TPC gas, we require the radial location of the first TPC point to be less than 70 cm. 
This cut causes an inefficiency of $12.0\pm2.0\%$ for non-photonic
electrons according to the estimates from both simulation and data.

Through embedding simulated electrons into high-tower trigger events and processing them through the same  software used for data production, single electron reconstruction efficiency in the TPC is found to be
0.84$\pm$0.04 with little dependence on momentum for $p_T>$ 2.0 GeV/$c$.

The analysis of the Run2005 data is slightly different from that of the Run2008 described above. Only half of the BEMC (0 $ < \eta <$ 1.0) was instrumented in 2005. Due to the presence of the silicon detectors, and their
significant material budget, photonic backgrounds were substantially higher. 
We select only tracks with $0 < \eta < 0.5$ from -30 cm 
$< V_z <$ 20~cm in order to avoid the supporting cone for the silicon detectors in the fiducial volume  while keeping track quality cuts identical to those in the Run2008 analysis.  However, we apply a tighter cut on the normalized ionization energy loss, i.e. $-0.7 < n\sigma_e < 3.0$, to improve hadron rejection.  BEMC clusters are grouped with geometrically overlapping BSMD clusters to
improve position resolution and electron hadron discrimination through
shower profile. The clustering algorithm is also modified to increase the 
efficiency of differentiating two overlapping BSMD
clusters by lowering the energy threshold of the second cluster
~\cite{bemc_ucla_cluster}. The minimum angle between track projection point at
the BEMC and all BEMC clusters is required to be less than $0.05$ radian.  We
also require each track to have more than one associated BSMD strip
in both $\phi$ and $\eta$ planes, and a tightened $p/E$ cut of $0.3 < p/E < 
1.5$, where $E$ is the energy of the associated BEMC cluster.  The
efficiencies for the electron identification cuts are estimated
by embedding simulated single electrons into minimum-bias PYTHIA~\cite{pythia_doc}
events.  Figure~\ref{fig:eID_eff} (b) shows the breakdown of 
electron identification efficiency as a function of $p_T$ in the
Run2005 analysis. There is no drop at high $p_T$ as in the 
Run2008 result because the energy of a whole BEMC cluster, instead of the highest tower,
is used for the $p/E$ cut.  No cut on the first TPC point is applied in this analysis.
To avoid the TPC tracking resolution effect that causes the broad shoulder extending toward higher masses in the invariant mass spectrum of the Run2008 analysis, 
we utilize a 2-D invariant mass by ignoring the opening angle in the $\phi$ plane 
when reconstructing the $e^+e^-$ invariant mass~\cite{bemc_ucla_cluster}. 
We require $-3 < n\sigma_{e} < 3$ for partner tracks, 2-D $m_{e^{+}e^-} < 0.1 $ GeV/$c^2$ for pairs, 
a maximum 0.1/0.05 radian for the opening angle in the $\phi$/$\theta$
plane, 
and a maximum 1.0 cm DCA between two electron helices. A cut of $p_T > 0.3$ GeV/$c$
for partners is also applied so that the simulation can describe the data
well.

By following independent analysis procedures from two RHIC runs where the amount of material for photonic background is significantly different, we will be able to validate our approach for measuring non-photonic electron production.

\subsection{Purity Estimation}
\label{purity_est}
After applying all electron identification cuts, the inclusive sample of primary electrons is still contaminated with hadrons. To estimate the purity of electrons in the inclusive sample, we perform a
constrained fit on the charged track $n\sigma_e$ distributions in different
$p_T$ regions with three Gaussian functions representing the expected
distributions of $\pi^{\pm}$, $K^{\pm}+p^{\pm}$ and $e^{\pm}$. The purity is estimated from the fit. 

\begin{figure}[t]
    \centering
    \includegraphics[width=1.0\linewidth]{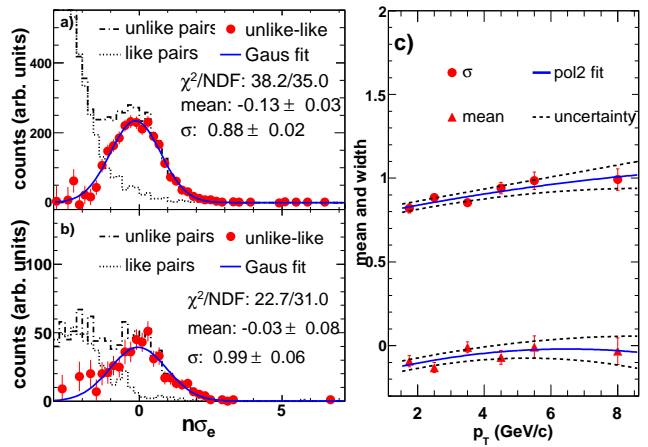}
    \caption{(Color online) Left two panels are $n\sigma_e$
        distributions in the Run2008 analysis for unlike-sign (dot-dashed line), 
like-sign (dotted line) and unlike-minus-like (closed circles) pairs together with a Gaussian fit (solid lines) at (a) 2.5 GeV/$c < p_T < 3.0$~GeV/$c$
        and (b) 8.0 GeV/$c < p_T < 10.0$~GeV/$c$ after
        applying all the electron identification cuts except the
        $n\sigma_e$ cut. Right panel (c) shows the mean and width of the 
        Gaussian fitting functions for pure photonic electron
        (unlike-minus-like) $n\sigma_e$ distribution as shown in left panels
        for each $p_T$ bin. See text for details.}
    \label{fig:nsig_pure_e}
\end{figure}

Ideally the  electron $n\sigma_{e}$ will follow the standard normal distribution. The actual distribution can be slightly different due to various effects in data calibrations. We can, however, determine its shape in different $p_T$ regions directly from data using  photonic electrons from the unlike-minus-like pairs.  The left panel of Fig.~\ref{fig:nsig_pure_e} shows the $n\sigma_e$ distribution for tracks with (a) 2.5 GeV/$c < p_T < 3.0$~GeV/$c$  and (b) 8.0 GeV/$c < p_T < 10.0$~GeV/$c$ from unlike-sign, like-sign pairs as well as for photonic electrons from the unlike-minus-like pairs. Here all electron identification cuts, except the $n\sigma_e$ cut, are applied. The $n\sigma_{e}$ of photonic electrons are well fitted with Gaussian functions.
Figure~\ref{fig:nsig_pure_e} (c) shows the mean and width of the Gaussian fit as a function of electron $p_T$, which, as discussed above, differ slightly from the ideal values. The solid lines in the figure are fits to the data using a second order polynomial function. 
The dotted lines are also second order polynomial fits to the data except that the data points are moved up and down simultaneously by one standard deviation.
The region between the dotted lines represents a conservative estimate of the fit uncertainty since we assume that the points are fully correlated.
The mean, width and their corresponding uncertainties from the fits are used to define the shape of electron $n\sigma_{e}$ distribution in the following 3-Gaussian fit. The $n\sigma_e$ of $\pi^{\pm}$ and $K^{\pm}+p^{\pm}$ are also expected to follow Gaussian
distributions~\cite{dedx}. Ideally their width is one and their means can be calculated through the Bichsel function~\cite{bichsel}. These ideal values are used as the initial values of the fit parameters in the following 3-Gaussian fit.  

\begin{figure}[t]
    \centering
    \includegraphics[width=1.0\linewidth]{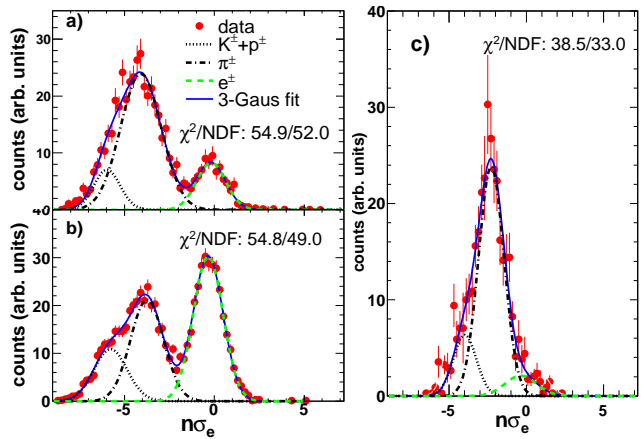}
    \caption{ (Color online) $n\sigma_e$ distribution for inclusive
        electrons (closed circles) and fits from
        different components at (a) 2.5 GeV/$c < p_T < 3.0$~GeV/$c$ in the Run2008 analysis, (b)  2.5 GeV/$c < p_T < 3.5$~GeV/$c$ in the Run2005 analysis and (c)
        8.0 GeV/$c < p_T < 10.0$~GeV/$c$ in the Run2008 analysis after applying all electron
        identification cuts except the $n\sigma_e$ cut. Different
        curves represent $K^{\pm}+p^{\pm}$ (dotted line),
        $\pi^{\pm}$ (dot-dashed line), electrons (dashed line) and the
        overall fit (solid lines)}.
    \label{fig:purity}
\end{figure}

Figure~\ref{fig:purity} shows the constrained 3-Gaussian fits to the
$n\sigma_e$ distributions of  inclusive electron candidates with 2.5 GeV/$c < p_T < 3.0$ GeV/$c$ in the Run2008 analysis (upper-left), 2.5 GeV/$c < p_T < 3.5$ GeV/$c$ in the Run2005 analysis (lower-left) and 8.0 GeV/$c < p_T < 10.0$~GeV/$c$ in the Run2008 analysis (right). Here we leave the $n\sigma_e$ cut open.  The dotted, dot-dashed and dashed lines represent, respectively, the fits for $K^{\pm}+p^{\pm}$,  $\pi^{\pm}$, and $e^{\pm}$. Compared to the Run2008 analysis, the electron component in the Run2005 analysis at similar $p_T$ is more prominent due to the larger conversion electron yield.  The solid lines are the overall fits to the spectra. 
The purity is calculated as the ratio of the integral of the electron
fit function to that of the overall fit function above the $n\sigma_e$ cut. 
No constraints are applied to the $K^{\pm}+p^{\pm}$ and $\pi^{\pm}$ functions unless the fits fail. 
To estimate the systematic uncertainty of the purity, the mean and width of the electron function are allowed
to vary up to one, two, three and four standard deviations from their central values. For each of the four constraints, we calculate one value of the purity. The final purity is taken as the mean and the systematic uncertainty is taken as the largest difference between the mean and the four values from the four constraints.   
To estimate the statistical uncertainty of the purity, we rely on a simple Monte-Carlo simulation. We first obtain a large sample of altered overall $n\sigma_e$ spectra by randomly shifting each data point in the original spectrum in Fig.~\ref{fig:purity} according to a Gaussian distribution with the mean and width set to be equal to the central value and the uncertainty of the original data point, respectively. We then obtain the purity distribution through calculating the purity from each of these altered spectra following the same procedure as discussed above. In the end, we fit the distribution with a Gaussian function and take its width as the statistical uncertainty. The total uncertainty of the purity is obtained as the quadratic sum of the statistical and systematic uncertainties.

\begin{figure}[t]
    \centering
    \includegraphics[width=1.0\linewidth]{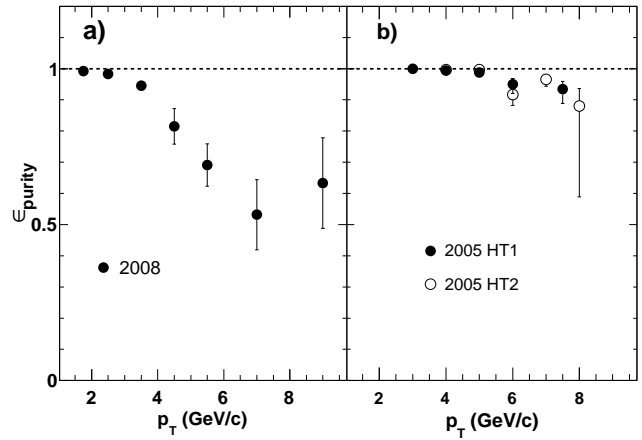}
    \caption{ Purity of the inclusive electron sample as a function of
        $p_T$ in data from (a) Run2008 and (b) Run2005. The
        2008 result is from combined datasets of all different high-tower
        triggers. The 2005 results for the two different high-tower
        triggers, i.e. HT1 (closed circles) and HT2 (open circles), are plotted separately.}
    \label{fig:purity05}
\end{figure}

We follow the same procedure in the Run2008 and the Run2005 analysis except that the overall $n\sigma_e$ distribution in the Run2008 analysis is the combined result from the datasets of all three high-tower triggers as described Sec.~\ref{Triggers}, while in the Run2005 analysis, the purity are calculated separately for the two high-tower triggers. Figure~\ref{fig:purity05} shows the purity as a function of electron $p_T$ for the Run2008 (a) and the Run2005 (b) data. Tighter electron identification cuts and much higher photonic electron yield lead to much higher purity for the Run2005 inclusive electron sample. 

\subsection{Photonic Electron Reconstruction Efficiency}
\label{pho_reco_eff}
Since photon conversions, $\pi^0$ and $\eta$ meson Dalitz decays are the dominant sources of photonic electrons, they are the components that we used to calculate $\epsilon_{pho}$, the photonic electron reconstruction efficiency,  in the analysis of the Run2008 data. The $\epsilon_{pho}$ for each individual component is calculated separately to account for its possible dependence on the decay kinematics of the parent particles. The final $\epsilon_{pho}$ is obtained by combining results from all components according to their relative contribution to the photonic electron yield. 

\begin{figure}[t]
    \centering
    \includegraphics[width=1.0\linewidth]{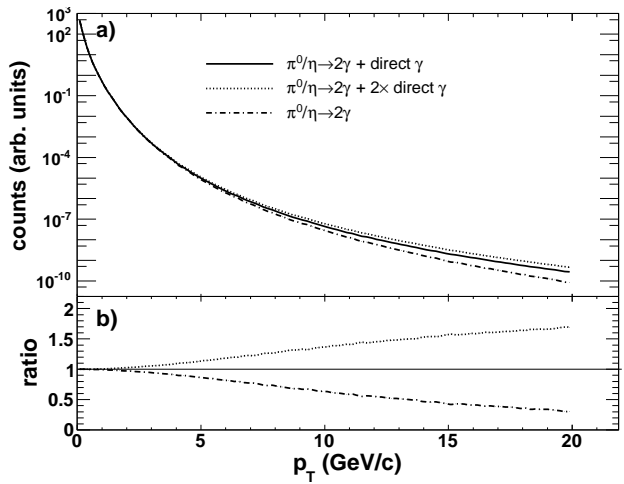}
    \caption{
        (a) Derived $p_T$ spectrum for inclusive photons (solid line) 
	and the uncertainty represented by the region between 
        the spectra of $\pi^0$ and $\eta$ decay photons (dot-dashed line) and
        inclusive photon with doubled direct photon yield (dotted
        line) as well as their ratio to the inclusive photon as shown in (b).}
    \label{fig:pi0_deca_gamma_dir}
\end{figure}

The determination of $\epsilon_{pho}$ is done through reconstructing electrons from simulated $\gamma$ conversion or Dalitz decay of $\pi^0$ and $\eta$ with uniform $p_T$ 
distributions that are embedded into high-tower trigger events. These events are 
then fully reconstructed using the same software chain as used for data analysis. 
To account for the efficiency dependence on the parent particle $p_T$, 
we use a fit function to the measured $\pi^0$ spectrum, 
the derived $\eta$ and inclusive photon $p_T$ spectra as weights. The fit function to the measured $\pi^0$ spectrum is provided by the PHENIX experiment in Ref.~\cite{phnx_dir_gamma}. The $\eta$ spectrum is derived from the $\pi^0$ measurement assuming $m_T$ scaling, i.e. replacing $p_T$ with $\sqrt{p^2_T + m^2_h - m^2_{\pi^0}}$ while keeping the function form unchanged.
Figure~\ref{fig:pi0_deca_gamma_dir} (a) shows the
derived inclusive $\gamma$ $p_T$ spectrum (solid line), and an estimate of its 
uncertainty represented by the region between the dotted and dot-dashed lines. Figure~\ref{fig:pi0_deca_gamma_dir} (b) shows the uncertainty in linear scale.
The inclusive $\gamma$ spectrum is obtained by adding the direct $\gamma$ yield 
to the $\pi^0$ and $\eta$ decay $\gamma$ yield calculated using PYTHIA. 
The direct $\gamma$ yield is obtained from the fit function to the direct $\gamma$ measurement provided by the PHENIX experiment in Ref.~\cite{phnx_dir_gamma}.
The dot-dashed line represents the $\gamma$ spectrum from $\pi^0$ and $\eta$  decay alone. The dotted line is obtained by doubling the direct $\gamma$ component in the inclusive photon spectrum. By comparing the ratio of the derived inclusive $\gamma$ yield to that of $\pi^0$ and $\eta$  decay photon with the double ratio measurement in Au+Au most peripheral collisions~\cite{phnx_dir_g_AuAu}, we found the uncertainty covers the possible variations of the inclusive photon yield.  

STAR simulations for $\gamma$ conversion and Dalitz decay are based on 
GEANT3 ~\cite{Brun:1987ma} which incorrectly treats Dalitz decays as 
simple 3-body decays in phase space. We therefore modified the GEANT
decay routines using the correct Kroll-Wada decay formalism
~\cite{Kroll:1955zu}. Their kinematics is strongly modified by the dynamic
electromagnetic structure arising at the vertex of the transition
which is formally described by a form factor. We included
the most recent form factors using a linear approximation for the
$\pi^0$ Dalitz decay ~\cite{Amsler:2008zzb}, and a pole approximation
for the decays of $\eta$ ~\cite{:2009wb}.

\begin{figure}[t]
    \centering
    \includegraphics[width=1.0\linewidth]{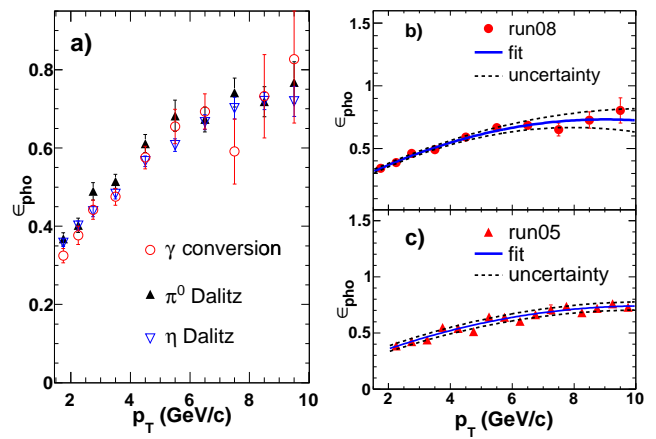}
    \caption{ (Color online) Photonic electron reconstruction efficiency as a
        function of $p_T$ for (a) $\gamma$ conversion (open circles), $\pi^0$ (closed triangles) and $\eta$ (open triangles) Dalitz decay for the Run2008
        analysis, (b) combination of $\gamma$ conversion, $\pi^0$ and $\eta$ 
        Dalitz decay for the Run2008 analysis and (c) $\gamma$ conversion
        for the Run2005 analysis. The solid line is a fit and the dashed lines represent the uncertainty. See text for details.}.
    \label{fig:pho_reco_eff}
\end{figure}

Figure~\ref{fig:pho_reco_eff} (a) shows the photonic electron reconstruction
efficiency as a function of electron $p_T$ for $\gamma$ conversion,  $\pi^0$
 and $\eta$ Dalitz decay electrons, which turn out to be very similar because of the similar decay kinematics. The increase towards larger electron $p_T$ is due to the higher probability of reconstructing both electrons from high $p_T$ (virtual) photons. 
The uncertainties shown in the plots are dominated by the statistics of the simulated events. The effect due to the variation of the inclusive
photon spectrum shape is found to be negligible for this analysis. Figure~\ref{fig:pho_reco_eff} (b) shows the combined photonic electron reconstruction efficiency for the Run2008 analysis, which is calculated as
\begin{eqnarray*}
    \epsilon_{pho} (p_T) = 
    & & \frac{N_{e}^{\gamma} (p_T)}{N_{e}^{\gamma} (p_T)+N_{e}^{\pi^0} (p_T)+N_{e}^{\eta} (p_T)} \cdotp \epsilon_{\gamma} (p_T) + \\ 
    & & \frac{N_{e}^{\pi^0} (p_T)}{N_{e}^{\gamma} (p_T)+N_{e}^{\pi^0} (p_T)+N_{e}^{\eta} (p_T)} \cdotp \epsilon_{\pi^0} (p_T) +
\\
    & & \frac{N_{e}^{\eta} (p_T)}{N_{e}^{\gamma} (p_T)+N_{e}^{\pi^0} (p_T)+N_{e}^{\eta} (p_T)} \cdotp \epsilon_{\eta} (p_T), 
\end{eqnarray*}
where $N_{e}^{\gamma}$, $N_{e}^{\pi^0}$ and $N_{e}^{\eta}$ are respectively the yield of electrons from photon conversion, $\pi^0$ and $\eta$ Dalitz decay;  $\epsilon_{\gamma}$,  $\epsilon_{\pi^0}$ and $\epsilon_{\eta}$ are the corresponding photonic electron reconstruction efficiencies. Based on Table~\ref{tab:material_run08},  approximately $36\%$ of the photonic electrons are from $\pi^0$ Dalitz decay and about $10\%$ are from $\eta$ Dalitz decay. 
Their variations have negligible effect on the results since $\epsilon_{\gamma}$,  $\epsilon_{\pi^0}$ and $\epsilon_{\eta}$  are almost identical. 
The solid line is a second order
 polynomial fit to the data. The systematic uncertainty is represented by the 
region between the dotted lines, which are second order
 polynomial fits after moving all the data points simultaneously up and down by one standard deviation.

For the Run2005 analysis, the dominant source of photonic electrons is conversion in the silicon detectors. 
We therefore neglect contributions from Dalitz decays while following the same procedure as for the Run2008 analysis to calculate $\epsilon_{pho}$. Figure~\ref{fig:pho_reco_eff} (c) shows $\epsilon_{pho}$ as a function of $p_T$ for
$\gamma$ conversion for the Run2005 analysis. The solid line is a fit to the spectrum with a second order polynomial function and the region between dashed lines represents the uncertainty estimated in the same way as for the Run2008 analysis. The inclusion of the Dalitz decays is estimated to reduce the $\epsilon_{pho}$ by less than $0.5\%$ which is well within the systematic error.  The uncertainty because of the inaccurate material distribution in the simulation as mentioned in Sec.~\ref{material_budget} is negligible since the majority of the material, dominated by our silicon detectors, is within a distance of 30 cm from beam pipe and the variation of $\epsilon_{pho}$ of photonic electrons produced within this region is small.

\subsection{Trigger Efficiency}
\label{trg_eff}

The trigger efficiency is the ratio of the electron yield from high-tower trigger events to that from minimum-bias trigger events after normalizing the two according to the integrated luminosity.
To have a good understanding of trigger efficiency, one needs enough minimum-bias events for the baseline reference. However, for the Run2008 $p+p$ data, the number of  minimum-bias events is too small to be used for this purpose. Fortunately the Run2008 {\it d}+Au data were taken using the same sets of high-tower triggers as used for $p+p$ run. Since the two data sets were taken serially, the high-tower trigger efficiency is expected to be the same. During the {\it d}+Au run, many events also were taken using the VPD trigger, which is essentially a less efficient minimum-bias trigger that can serve as the baseline reference for trigger efficiency analysis.  As a cross check, we also evaluate the trigger efficiency through the Run2008 $p+p$ simulation.

\begin{figure}[t]
    \centering
    \includegraphics[width=1.0\linewidth]{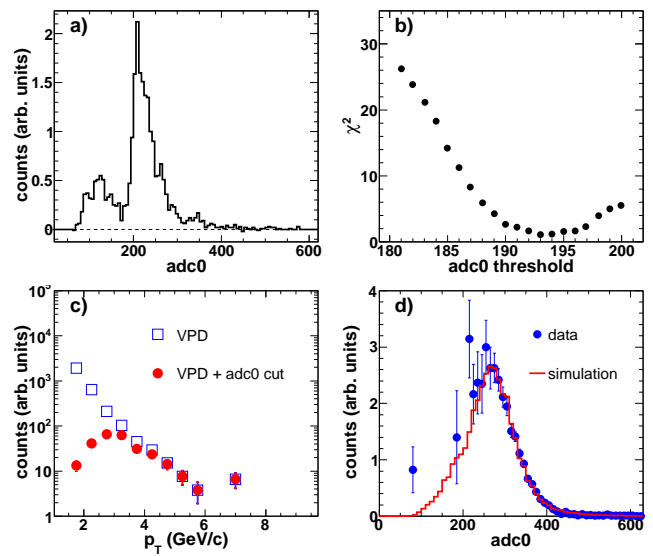}
    \caption{(Color online) (a) $adc0$ distribution for high-tower
        trigger events. (b) $\chi^2$ as a function of the $adc0$ cut. (c)
        Raw inclusive electron $p_T$ spectrum from VPD trigger in
        Run2008 {\it d}+Au collisions before (open squares) and after applying
        the $adc0$ $>$ 193 cut (closed circles). (d) $adc0$
        distribution for data (closed circles) and simulation (solid
        line) at $p_T = 4.0-5.0$ GeV/$c$.  See text for details.}
    \label{fig:adc0_chi2_sim_data_VPD}
\end{figure}

From the VPD trigger events, we first regenerate a high-tower trigger $p_T$ spectrum by requiring $adc0$ of BEMC clusters to be larger than the threshold. The $adc0$ is the offline ADC value of a BEMC cluster's most energetic tower which is one of the high-towers responsible for firing a high-tower trigger. Figure~\ref{fig:adc0_chi2_sim_data_VPD} (a) shows the $adc0$ distribution of photonic electrons from high-tower trigger events. The sharp cut-off around a value of 200 is the offline ADC value of the trigger threshold setting. The smaller peak below the trigger threshold is due to electrons which happen to be in events triggered by something else other than the electrons. By requiring $adc0$ to be larger than the threshold,  we reject these electrons which did not trigger the event since the uncertainty of their yield is affected by many sources and  is therefore hard to be evaluated reliably. When the threshold is correctly chosen, the regenerated spectrum shape should be very similar to that of the actual high-tower trigger. We therefore quantitatively determine the trigger threshold as the $adc0$ cut which minimizes the $$\chi^2 =
\displaystyle\sum_{i}(\frac{N_i(VPD+adc0)}{N_i(HT)}-1)^2/\sigma_{i}^2$$
where $N_i(VPD+adc0)$ is the regenerated high-tower trigger electron yield from VPD
events in the $i^{th}$ $p_T$ bin, $N_i(HT)$ is the electron yield at the
same $p_T$ bin from the actual high-tower trigger events, and $\sigma_i$ is the
uncertainty of $N_i(VPD+adc0)$. Figure~\ref{fig:adc0_chi2_sim_data_VPD} (b) shows the $\chi^2$
as a function of the $adc0$ cut; the threshold is taken as 193. Figure~\ref{fig:adc0_chi2_sim_data_VPD} (c) shows the  $p_T$ spectrum of raw inclusive electrons from the VPD trigger (open squares) and the regenerated high-tower spectrum (closed circles) after applying the $adc0$ $>$ 193 cut used to calculate the trigger efficiency. 

To estimate trigger efficiency through simulation, we tune the simulated single electron $adc0$ spectrum in each individual $p_T$ bin to agree with the data in the region above the threshold. The data spectra are obtained from the unlike-minus-like pairs, i.e. pure photonic electrons. As a demonstration of the comparison, Fig.~\ref{fig:adc0_chi2_sim_data_VPD} (d) shows the spectra from data (closed circles) and simulation  (solid line) at 4.0 GeV/$c < p_T < 5.0$ GeV/$c$. The efficiency is defined as the fraction of the simulated $adc0$ spectrum integral above the trigger threshold.

\begin{figure}[t]
    \centering
    \includegraphics[width=1.0\linewidth]{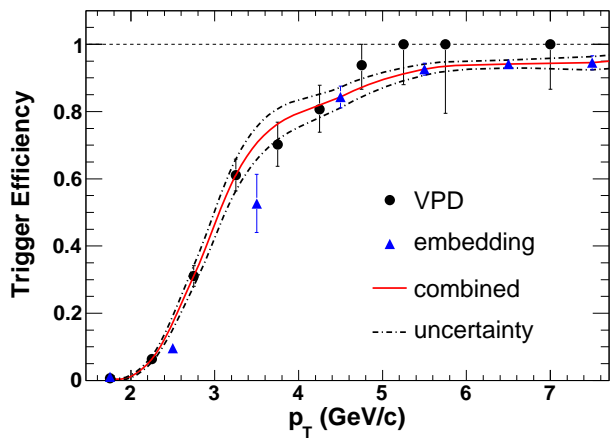}
    \caption{(Color online) $p_T$ dependence of high-tower trigger
        efficiency from data (closed circles), simulation (closed
        triangles) and combined results (solid line) for Run2008 analysis. 
        The dashed lines
        represent the uncertainty. See text for details.}
    \label{fig:trg_eff}
\end{figure}

In the Run2008 analysis the raw $p_T$ spectrum of non-photonic electrons is obtained by combining the datasets of all three high-tower triggers. Since the shape of the combined spectrum is the same as that of the high-tower trigger with the lowest threshold, we only need to estimate the trigger efficiency of this lowest threshold trigger. Figure~\ref{fig:trg_eff} shows the trigger efficiency as
a function of $p_T$ that is calculated using {\it d}+Au VPD events (closed circles) and simulated events (closed triangles) in the Run2008 analysis. At $p_T > 3.5$ GeV/$c$, they agree with each other reasonably well.
At lower $p_T$, the simulated results are not reliable because the
numerator in the efficiency calculation is only from a tail of the
spectrum and a small mismatch between simulation and data can have a
large impact on the results. On the other hand, the results from VPD events suffer from low statistics at high $p_T$. The final efficiency is therefore taken to be the combination of the two, i.e., at $p_T < 3.5$ GeV/$c$, the efficiency is equal to that from VPD events assigning a systematic uncertainty identical to the statistical uncertainty of the data point, while at high $p_T$, the efficiency is equal to the simulated results, and the systematic error are from the tuning uncertainty. 

\begin{figure}[t]
    \centering
    \includegraphics[width=1.0\linewidth]{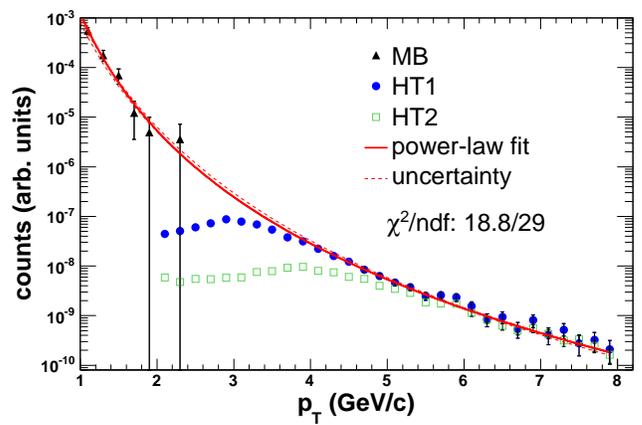}
    \caption{(Color online) Raw inclusive electron $p_T$ spectrum for
        minimum-bias (closed triangles) and two high-tower triggers, i.e.
        HT1 (closed circles) and HT2 (open squares) together with a
        power-law fit (solid line) and fit uncertainty (dashed line) for Run2005 analysis.}
    \label{fig:combine05}
\end{figure}

\begin{figure}[t]
    \centering
    \includegraphics[width=1.0\linewidth]{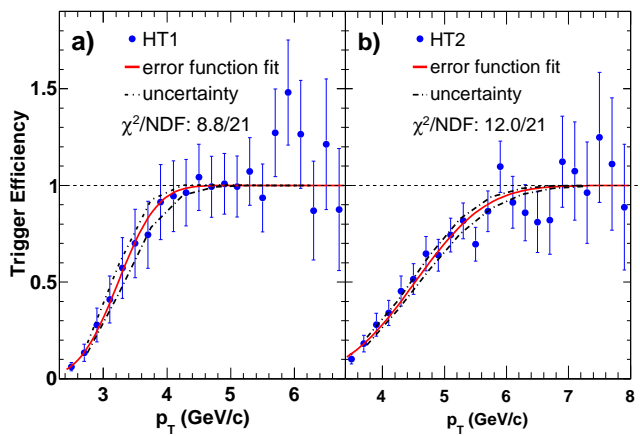}
    \caption{(Color online) $p_T$ dependence of trigger efficiency for
        the two high-tower triggers, i.e. HT1 (a) and HT2 (b)
        including result from data (closed circles) for Run2005, an error function
        fit (solid lines) and the fit uncertainty (dot-dashed lines).}
    \label{fig:trg_eff05}
\end{figure}

In the Run2005 analysis, the efficiencies of the two high-tower triggers are estimated separately. While there are more minimum-bias events  for Run2005 than for Run2008, the statistics are poor at $p_T > 2.0$ GeV/$c$. We thus rely on a fit to the spectrum, which consists of  minimum-bias events at low $p_T$ and high-tower trigger events at high $p_T$ where the trigger is expected to be fully efficient, as the baseline reference for the trigger efficiency evaluation. Figure~\ref{fig:combine05} shows the raw inclusive electron $p_T$ spectrum from minimum-bias, HT1 and HT2 events.  The fit uses a
power-law function $A(1+p_T/B)^{-n}$.  The regions where we expect HT1 and HT2 trigger to be fully efficient are above $4.5$ and $6.0$ GeV/$c$ respectively.  The dashed line shows the fit
uncertainty, obtained from many fit trials.  In a single fit trial,
each data point is randomized with a Gaussian random number, with the
mean to be the central value and the rms to be the statistical uncertainty
of the data point.  Additional systematic uncertainty coming from  fits using different functions is included in the cross section calculation and is not displayed in the figure.  Figure~\ref{fig:trg_eff05} shows the efficiency of HT1 (a) and HT2 (b) triggers, defined as the ratio of the raw HT1 or HT2 inclusive electron spectrum to the baseline fit function.  We used error functions to parameterize both efficiencies. The dashed lines represent the uncertainty, obtained in the same way as for the Fig.~\ref{fig:combine05} fits.

High-tower trigger efficiency for photonic and non-photonic electrons can be different. Unlike non-photonic electrons, a photonic electron always has a partner. In case both
share the same BEMC tower, the deposited energy will be higher than that for an
isolated electron and will lead to a higher efficiency. 
The effect can be  quantified by  comparing the ratio of the isolated electron yield to the yield of  electrons  with partners in minimum-bias events to the same ratio in high-tower trigger events.
We found that the difference is negligible at $p_T > 2.5 $ GeV/$c$, while the trigger efficiency for photonic electron is $~20-30\%$ higher than for non-photonic electron in the lower $p_T$ region.

\subsection{Stability of the Luminosity Monitor}
\label{lum}
\begin{figure}[t]
    \centering
    \includegraphics[width=1.0\linewidth]{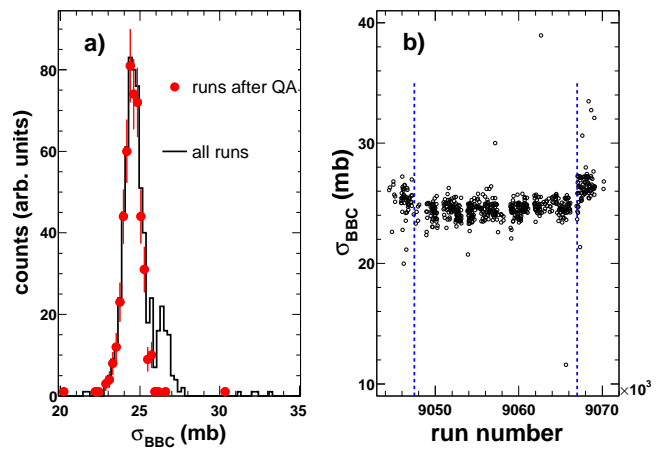}
    \caption{(Color online) (a) Distribution of the calculated
        $\sigma_{BBC}$ before (solid lines) and after (closed circles)
        removing events at the beginning and the end of Run2008. 
        (b) Variation of the calculated $\sigma_{BBC}$ as a
        function of run number. The runs outside the region between the two dashed lines are rejected.}
    \label{fig:HT2_lum}
\end{figure}

The BBC trigger was used to monitor the integrated
luminosity for Run2005. During Run2008, because of the large
beam background firing the BBC trigger, a high threshold high-tower
trigger was used as the luminosity monitor. To quantify the stability of the
monitor with respect to BBC, we calculate the BBC cross section as a
function of run number using $\sigma_{BBC} = N_{minbias}/\cal L$,
where $\cal L$ $ = N_{mon}/ \sigma_{mon}$, $\sigma_{mon}$ is the monitor cross section which is estimated to be $1.49$ $\mu$b using low luminosity runs, $N_{minbias}$ and $N_{mon}$
are respectively the number of events from the BBC trigger and the monitor 
after correcting for prescaling during data acquisition. 
Here a run refers to a block of short term ($\sim$30 minutes) data taking.
Figure~\ref{fig:HT2_lum} (a) shows the distribution of the calculated
$\sigma_{BBC}$. There are two peaks in the figure. The dominant one
centered around 25 mb contains most of the recorded luminosity in Run2008. 
The minor one centered at a higher value comes from events taken at
the beginning and the end of Run2008 represented by the regions beyond the two dashed lines in  
Fig.~\ref{fig:HT2_lum} (b) showing the calculated BBC cross section as a function of run number. After removing these runs 
taken at the beginning and the end of Run2008, the minor peak disappeared and  
the performance of the monitor appeared to be very stable. 

In the data analysis, we also reject those with
$\sigma_{BBC} < 20$ mb or $\sigma_{BBC} > 30$ mb. We fit the
$\sigma_{BBC}$ distribution with a Gaussian function and assign the
width of the function ($2.3\%$) as the systematic uncertainty of the
$\sigma_{mon}$ with respect to BBC cross section.

The integrated luminosity sampled by the high-tower triggers are $\sim$2.6
pb$^{-1}$ and $\sim$2.8 pb$^{-1}$ for Run2008 and Run2005, respectively.

\subsection{Contribution from Vector Mesons}
\label{feeddown}

The main background sources of electrons that do not originate from photon conversion and Dalitz decay are
electromagnetic decays of heavy ($J/\psi$, $\Upsilon$) and light vector mesons ($\rho$, $\omega$ and $\phi$) as well as those from Drell-Yan process.

\begin{figure}[t]
    \centering
    \includegraphics[width=1.0\linewidth]{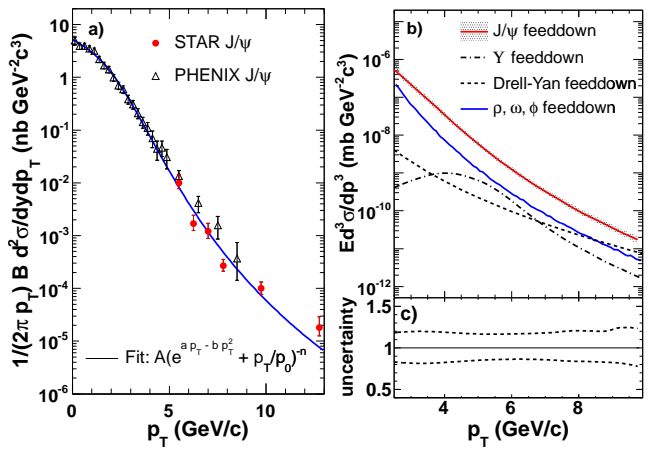}
    \caption{(Color online) (a) The $J/\psi$ invariant 
cross section measurement from STAR (closed circles) and PHENIX (open triangles), together with the fits using $A(\exp(a p_{T} - b p_{T}^{2}) + p_{T}/p_{0})^{-n}$ (solid line).   
        (b) Invariant cross section of the electron from decays of $J/\psi$ (band), $\Upsilon$ (dot-dashed line), Drell-Yan (dotted line)   
and light vector mesons (solid line). The uncertainty of the $J/\psi$ feed-down is represented by the band shown in (c) in linear scale.
        }
    \label{fig:feeddown}
\end{figure}

The electrons from \jpsi\ decay contribute noticeably to the observed 
non-photonic electron signal as
pointed out in Ref.~\cite{Adare:2010de}.  
In order to estimate the contribution from $\jpsi \rightarrow \epem$
to the non-photonic electron yield, we combine the measured
differential \jpsi\ cross-sections from PHENIX ~\cite{Adare:2006kf} and
STAR ~\cite{Abelev:2009qaa}. For each data point we add the statistical and systematic 
uncertainties, except the global uncertainties, in quadrature.
Figure~\ref{fig:feeddown} (a) shows the measured $J/\psi$ differential cross section 
from the two experiments.
While the PHENIX measurement dominates the low to
medium-\pT\ region, the STAR measurement dominates the high-$p_T$ region. 
The combined spectrum is fit using a power-law function of the form $E
d^3\sigma/d^3p|_{y=0} = A (\exp(a \pT - b \pT^2) + p_T/p_0)^{-n}$
where $A$ = 5.24$\pm$0.87 mb$\cdot {\rm GeV^{-2}}c^3$, $a$ = 0.32$\pm$0.04 ${\rm GeV^{-1}}c$, $b$ = 0.06$\pm$0.03 ${\rm GeV^{-2}}c^2$, $p_0$ = 2.59$\pm$0.21 GeV/$c$ and $n$ = 8.44$\pm$0.61 are fit parameters. The $\chi^2$/NDF of the fit is  27.8/25. To obtain the uncertainty of the fit, the global uncertainties of the STAR and the PHENIX (10\%~\cite{cesar}) measurements are assumed to be uncorrelated.  We move the PHENIX data up by 10\% and repeat the fit to obtain the band of 68\% confidence intervals. The upper edge of the band is treated as the upper bound of the fit. Following the same procedure except moving the PHENIX data down by 10\%, we obtain the lower bound of the fit as the lower edge of the band.
Furthermore, since we are considering a rather
large \pT\ range ($p_T < 14.0$ \gevc\ ),  we cannot assume that the \pT\ and
rapidity distributions factorize. 
We use PYTHIA to generate $dN/dy(p_T)$ and implement a Monte-Carlo program
using the above functions as probability density functions
to generate \jpsi\ and decay them into \epem\ assuming the \jpsi\ to
be unpolarized. The decay electrons are filtered through the same
detector acceptance as used for the non-photonic electrons. The band in Fig.~\ref{fig:feeddown} (b) shows the invariant cross section of $J/\psi$ decay electrons as a function of the electron $p_T$. The uncertainty of the derived yield comes from the uncertainty of the fit to the $J/\psi$ spectra and is represented by the band which is also shown in Fig.~\ref{fig:feeddown} (c) in linear scale.  

The invariant cross section of electrons from $\Upsilon$ decays ($\Upsilon \rightarrow e^+ e^-$), represented by the dot-dashed line in Fig.~\ref{fig:feeddown} (b), is calculated in a similar fashion as that for the $J/\psi$ except that the input $\Upsilon$ spectrum is from a Next-to-Leading Order
pQCD calculation in the color evaporation model (CEM)~\cite{vogt_upslion}.
We have to rely on model calculations since so far no invariant
$p_T$ spectrum in our energy range has been measured. However, in a recent measurement STAR reported the overall production cross section
for the sum of all three $\Upsilon$(1S+2S+3S) states in $p$+$p$ collisions at $\sqrt{s}$ = 200 GeV to be $B \times d\sigma/dy = 114\pm38^{+23}_{-24}$ pb, which is consistent with the CEM prediction~\cite{Abelev:2010am}. Adding the statistical and systematic uncertainty in quadrature, the total relative uncertainty of this measurement is $\sim$39\%, which is the value we assigned as the total uncertainty of the $\Upsilon$ feed-down contribution
to the non-photonic electrons at all $p_T$.

The contribution to the non-photonic electron yield from the light vector mesons
is estimated using PYTHIA, assuming  the meson spectra follow $m_T$ scaling. 
We generate a sample of decay electrons using light vector mesons with flat spectra in $p_T$ as input.
To derive the differential cross section of the electrons, we 
keep only those electrons within the same detector acceptance as that for the non-photonic electrons and weight them with 
the spectra of  $\rho$, $\omega$ and $\phi$. The meson spectra are obtained by replacing the $p_T$ 
with $\sqrt{p^2_T + m^2_h - m^2_{\pi^0}}$ 
in the same fit function as for the $\pi^0$ measurement 
(see Fig.~\ref{fig:pi0_deca_gamma_dir}). Here $m_h$ is the mass of the vector meson.
The relative yields of the mesons to $\pi$~\cite{Adare:2006hc} are also taken into account during this process. We include the decay channels 
$\phi \rightarrow e^+e^-$, $ \phi\rightarrow \eta e^+e^-$, $\omega\rightarrow e^+e^-$, $\omega\rightarrow \pi^0 e^+e^-$ and $\rho\rightarrow e^+e^-$ in the calculation.
The derived electron differential cross section is represented by the  solid line in Fig.~\ref{fig:feeddown} (b). We assign a 50$\%$ systematic uncertainty to cover the uncertainty of 
the $\pi^0$ measurement and the meson to pion ratios. 

The contribution to the non-photonic electron yield from the Drell-Yan processes is represented by the dotted line in Fig.~\ref{fig:feeddown} (b) and 
is estimated from a Leading-Order pQCD calculation using the CTEQ6M parton
distribution function with a K-factor of 1.5 applied and without a cut on the electron pair mass~\cite{werner}. No uncertainty is assigned to this estimate.

\section{Results}
\label{result}
\subsection{Non-photonic Electron Invariant Cross Section}
\label{inv_cs}
\begin{figure}[b]
    \centering
    \includegraphics[width=1.0\linewidth]{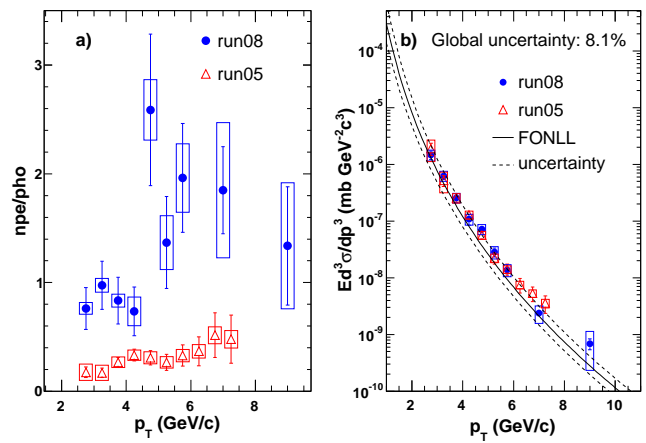}
    \caption{(Color online) (a) Ratio of non-photonic to photonic
        electron yield from the Run2008 (closed circles) and the Run2005 (open
        triangles) analyses. (b) Invariant cross section for non-photonic
        electron production ($\frac{e^++e^-}{2}$) in $p+p$ collisions
        from the Run2008 (closed circles) and the Run2005 (open
        triangles) analyses. The error bars and the boxes represent statistical and systematic uncertainty, respectively. The solid line is FONLL calculation and the dashed lines are the
        FONLL uncertainties~\cite{Cacciari:2005rk}. }
    \label{fig:nph_over_ph_ratio}
\end{figure}

\begin{figure}[t]
    \centering
    \includegraphics[width=1.0\linewidth]{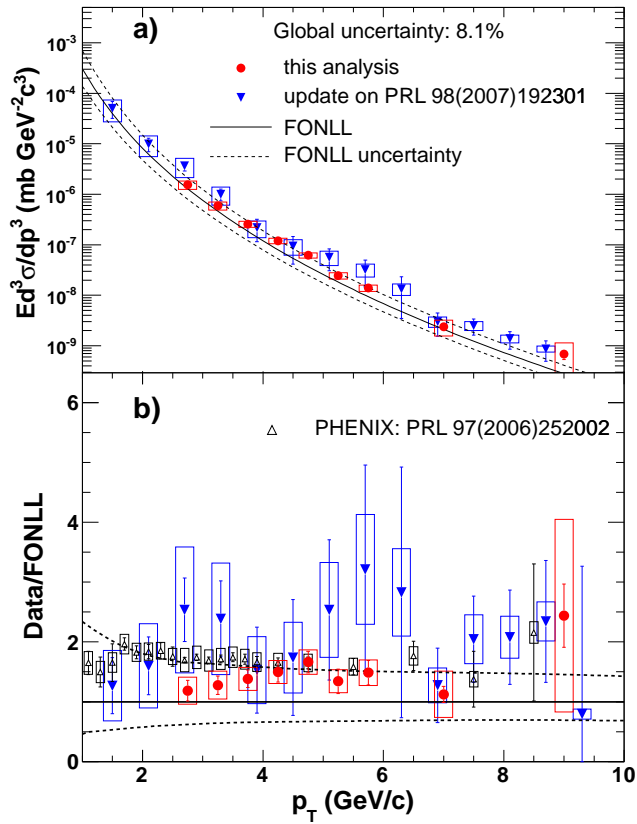}
    \caption{(Color online) (a) Invariant cross section of
        non-photonic electron production ($\frac{e^++e^-}{2}$) in
        $p+p$ collisions from this analysis (closed circles) after combining results from Run2005
        and Run2008. The published STAR result~\cite{Abelev:2006db} (closed triangles) is also shown. (b) Ratio of data over FONLL~\cite{Cacciari:2005rk} from all measurements at RHIC including PHENIX results~\cite{Adare:2006hc} (open
        triangles).}
    \label{fig:inv_cross_sect_combine}
\end{figure}

The invariant cross section for non-photonic electron production is calculated according to 

\begin{displaymath}
    E
    \frac{d^3\sigma}{dp^3} \,=\,
    \frac{1} {\cal L} \, 
    \frac{1} { {2\pi} \, {p_T} \, {\Delta}p_T \,{\Delta}y} \,
    \frac{N_{npe}} {{\epsilon_{rec}} \, {\epsilon_{trig}} \, { \epsilon_{eid}} \, {\epsilon_{BBC}}} ,
    \label{eq:dsigdpt}
\end{displaymath}
where $N_{npe}$ is the non-photonic electron raw yield with the $V_z$ cuts,
$\epsilon_{rec}$ is the product of the single electron reconstruction efficiency and the correction factor for momentum resolution and finite spectrum bin width,
$\epsilon_{trig}$ is the high-tower trigger efficiency,
$\epsilon_{eid}$ is the electron identification efficiency,
$\cal L $ is the integrated luminosity with the $V_z$ cuts and 
$\epsilon_{BBC} = 0.866\pm0.08$ is the BBC trigger efficiency. 
The systematic uncertainties of all these quantities are listed in Table~\ref{tab:syst_error}. 
The relative uncertainty of $\cal L\cdot$$\epsilon_{BBC}$ in maximum range is 14\% including uncertainties in tracking efficiency~\cite{bbc_cross_section}. 
Assuming a flat distribution within the range, we estimate the $\cal L\cdot $$\epsilon_{BBC}$ uncertainty to be 8.1\% in one standard deviation. 
The uncertainty of $N_{npe}$ is the quadratic sum of the uncertainties from the estimation of $\epsilon_{pho}$, purity and the light vector meson contribution.  
The uncertainty of $\epsilon_{rec}$ is the quadratic sum of the uncertainties from correcting the track momentum resolution, the finite spectrum bin width as well as the estimation of single electron reconstruction efficiency.  
The range of the uncertainty for each individual quantity covers the
variation of the uncertainty as a function of $p_T$ .
In order to compare with the result in Ref.~\cite{Abelev:2006db, Adare:2006hc}, we do not subtract the $J/\psi$, $\Upsilon$ and Drell-Yan contribution from the non-photonic electron invariant cross section shown in Fig.~\ref{fig:nph_over_ph_ratio} and Fig.~\ref{fig:inv_cross_sect_combine}. 

\begin{table} [t]
    \caption{\label{tab:syst_error}
        Sources of systematic uncertainty for the non-photonic electron invariant yield in $p+p$ collisions. Type A are point to point uncertainties. Type B are scaling uncertainties which move data points in the same direction. Type C are the scaling uncertainties that are common to both Run2008 and Run2005. The range in each individual source covers the variation of the systematic uncertainty as a function of $p_T$.}
    \begin{ruledtabular} \begin{tabular}{|c|cc|}
            source 					  & Run2008 									& Run2005 \\
            \hline
            $N_{npe}$              & 5.0-48.1~\% (A) 					&  8.5-38.0~\% (A)\\
            \hline
            $\epsilon_{eid}$       & 6.5-25.2~\% (A)  						&0.7-2.0~\% (A) \\
            \hline
            $\epsilon_{trg}$       & 1.8-10.0~\% (A)  					&0.3-16~\% (A)\\
            							  & 5.4~\% (B) 								&		\\
            \hline
            $\epsilon_{rec}$       & 2.3-33.3~\% (A)   				   &  1.0-3.5~\% (A) \\
           								  & 15.7~\%  (B)							&	11.0~\% (B)	\\
            \hline
            $\cal L\cdot$$\epsilon_{BBC}$               & 2.3~\% (B)   						&   \\
                  											   & 8.1~\% (C)   						&  8.1~\% (C)\\
        \end{tabular} \end{ruledtabular}
\end{table}

Figure~\ref{fig:nph_over_ph_ratio} (a) shows the ratio of non-photonic
to photonic electron yield as a function of $p_T$ in $p+p$
collisions in Run2008 (closed circles) and Run2005 (open triangles). The ratio for Run2008 is
much larger because there was much less material in front
of the TPC for Run2008. Figure~\ref{fig:nph_over_ph_ratio} (b) shows the
non-photonic electron invariant cross section ($\frac{e^++e^-}{2}$) as
a function of $p_T$ in $p+p$ collisions from the Run2008  analysis (closed circles) and the Run2005 analysis (open triangles). Despite the large difference in photonic background, the two measurements are in good agreement.

Figure~\ref{fig:inv_cross_sect_combine} (a) shows 
the non-photonic electron ($\frac{e^++e^-}{2}$) invariant cross
section obtained by combining the Run2008 and the Run2005 results using the ``Best Linear Unbiased Estimate~\cite{blue}. The corrected result of our early 
published measurement using year 2003 data ~\cite{Abelev:2006db} 
is shown in the plot as well. 
The published result exceeded pQCD predictions from FONLL calculations 
by about a factor of four.  We, however, uncovered a mistake in the corresponding analysis in calculating $\epsilon_{pho}$. The details are described in the erratum~\cite{Abelev:2006db}. 
To see more clearly the comparison,
Fig.~\ref{fig:inv_cross_sect_combine} (b) shows the ratio of each
individual measurement, including PHENIX results, to the FONLL
calculation. One can see that all measurements at RHIC on non-photonic
electron production in $p+p$ collisions are now consistent with each
other. The corrected run 2003 data points have large uncertainties
because of the small integrated luminosity ($\sim$100 nb$^{-1}$) in that run. FONLL is able to describe the RHIC measurements within its theoretical uncertainties.

\subsection{Invariant Cross Section of Electrons from \\Bottom and Charm Meson Decays }
\label{b_quark}
Electrons from bottom and charm meson decays are the two dominant components of the 
non-photonic electrons. Mostly due to the decay kinematics, the azimuthal 
correlations between the daughter electron and daughter hadron are different 
for bottom meson decays and charm meson decays. A study of these azimuthal 
correlations has been carried out on STAR data and is compared with a PYTHIA 
simulation to obtain the ratio of the bottom electron yield to the heavy flavor decay electron yield ($e_B/(e_B + e_D)$)~\cite{e_h}, where PYTHIA was tuned to reproduce STAR measurements of D mesons $p_T$ spectra ~\cite{pythia_tune}.  Using the measured $e_{B}/(e_B + e_D)$, together with the measured non-photonic electron cross section with the electrons from $J/\psi$, $\Upsilon$ decay and Drell-Yan processes subtracted, we are able to disentangle these two components. 

\begin{figure*}[t]
\centering
\includegraphics[width=0.7\linewidth]{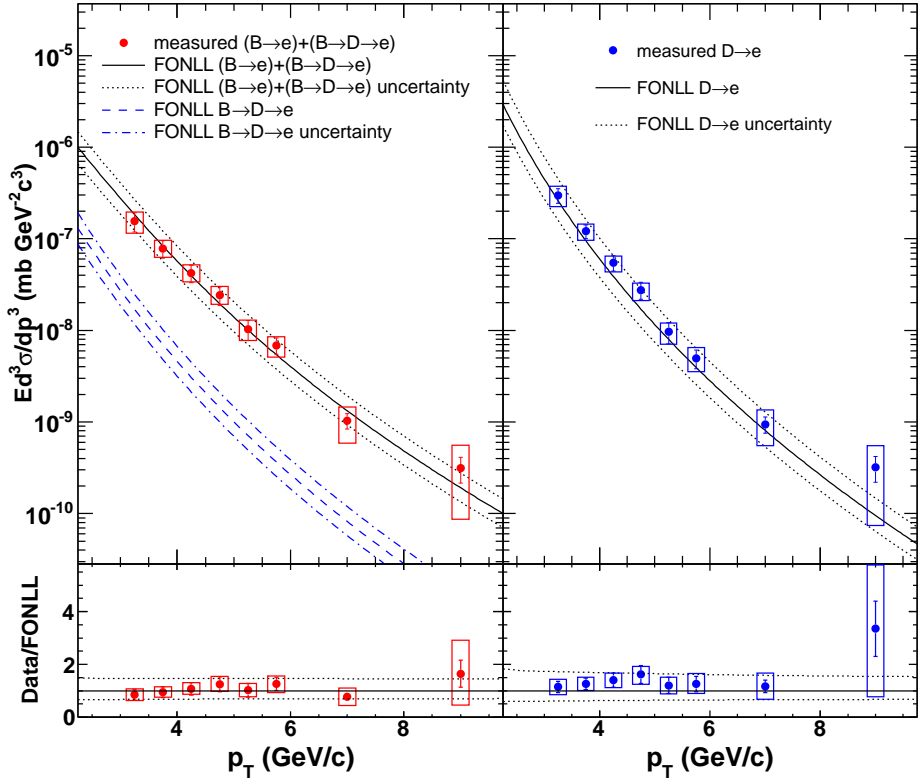}
\caption{(Color online) Invariant cross section of electrons ($\frac{e^++e^-}{2}$) from bottom (upper-left) and charm meson (upper-right) decay, together with the ratio of the corresponding measurements to the FONLL predictions for bottom (lower-left) and charm electrons (lower-right). The solid circles are experimental measurements.  The error bars and the boxes are respectively the statistical and systematic uncertainties. The solid and dotted curves are the FONLL predictions and their uncertainties. The dashed and dot-dashed curves are the FONLL prediction for B$\to$D$\to$e, i.e. electrons from the decays of D mesons which in turn come from B meson decays.} 
\label{fig:BandD}
\end{figure*}

The bottom electron cross section is calculated as $e_B/(e_B + e_D)$ times the non-photonic electron cross section with the contribution from $J/\psi$, $\Upsilon$ decay and Drell-Yan processes subtracted. The same procedure applies to the charm electrons except that ($1-e_{B}/(e_B + e_D))$ is used instead. The specific location of $p_T$ where the $e_B/(e_B + e_D)$ is measured, is different from that of the non-photonic electrons. To accommodate the difference, we calculate $e_B/(e_B + e_D)$ in any given $p_T$ in non-photonic electron measurements through a linear interpolation of the actual $e_B/(e_B + e_D)$ measurements. 
As an estimation of the systematic uncertainty of the interpolated value, we also repeat the same procedure using the curve predicted by FONLL. 
Figure ~\ref{fig:BandD} shows the invariant cross section of electrons ($\frac{e^++e^-}{2}$) from bottom (upper-left) and charm (upper-right) mesons as a function of $p_T$ and the corresponding FONLL predictions, along with the ratio of each measurement to the FONLL calculations (lower panels). The statistical uncertainty of each data point is obtained by adding the relative statistical uncertainties of the corresponding data points in the non-photonic electron and the $e_B/(e_B + e_D)$ measurement in quadrature. The systematic uncertainties are treated similarly, except that the  uncertainties from the interpolation process are also included. 
The measured bottom electrons are consistent with the central value of FONLL calculation and the charm electrons are in between the central value and upper limit of the FONLL calculation, the uncertainties of which are from the variation of heavy quark masses and scales.  From the measured spectrum, we determine the integrated cross section of  electrons  ($\frac{e^++e^-}{2}$) at ~3 GeV/$c < \pT < $ 10 GeV/$c$ from bottom and charm meson decays to be, respectively, $$ {d\sigma_{(B\to e)+(B\to D \to e)} \over dy_e}|_{y_e=0} = 4.0\pm0.5({\rm stat.})\pm1.1({\rm syst.}) {\rm nb} $$ $${d\sigma_{D\to e} \over dy_e}|_{y_e=0} = 6.2\pm0.7({\rm stat.})\pm1.5({\rm syst.}) {\rm nb},$$ where $y_e$ is the electron rapidity. The 8.1\% global scale uncertainty from the BBC cross section is included in the total systematic uncertainty.

Relying on theoretical model predictions to extrapolate the measured results to the phase space beyond the reach of the experiment, one  can estimate the total cross section for charm or bottom quark production. 
We perform a PYTHIA calculation with the same parameters as in Ref.~\cite{Adams:2004fc}. After normalizing the $p_T$ spectrum to our  high-\pT~ measurements and extrapolating the results to the full kinematic phase space, we obtain a total bottom production cross section of 1.34 $\mu$b. 
However, with the PYTHIA calculation using the same parameters except MSEL=5, i.e. bottom production processes instead of minimum-bias processes as in the former calculation, we obtain a value of 1.83 $ \mu$b.
The PYTHIA authors recommend the minimum-bias processes~\cite{pythia_doc}.
This large variation between the extracted total bottom production cross sections comes mostly from the large difference in the shape of the bottom electron spectrum in the two PYTHIA calculations with MSEL=1 and with MSEL=5. Since both calculations are normalized to the measured data, the difference in the shape shows up at $p_T$ $<$ 3 GeV/$c$. The fact that the PYTHIA calculation with MSEL=5 only includes leading order diagrams of bottom production causes the difference between the PYTHIA calculations. 
Measurements in the low $p_T$ region are therefore important for the understanding of bottom quark production at RHIC. Both values are consistent with the FONLL~\cite{Cacciari:2005rk} prediction, $1.87^{+0.99}_{-0.67}\mu$b, within its uncertainty.

\section{Conclusions}
\label{conclusion}

STAR measurements of high $p_T$ non-photonic electron production in $p+p$
collisions at $\sqrt{s}$= 200 GeV using data from Run2005 and Run2008 
agree with each other 
despite the large difference in background. This measurement and PHENIX measurement are consistent with each other within the quoted uncertainties. After correcting a mistake in the photonic electron reconstruction efficiency, the published STAR result using year 2003 data is consistent with our present measurements. We are able to disentangle the electrons from bottom and charm meson decays in the non-photonic electron spectrum using the measured ratio of $e_B/(e_B + e_D)$ and the measured non-photonic cross section.  The integrated bottom and charm electron cross sections  ($\frac{e^++e^-}{2}$) at 3 GeV/$c < \pT < $ 10 GeV/$c$ are determined separately as

$$ {d\sigma_{(B\to e)+(B\to D \to e)} \over dy_e}|_{y_e=0} = 4.0\pm0.5({\rm stat.})\pm1.1({\rm syst.}) {\rm nb} $$ $${d\sigma_{D\to e} \over dy_e}|_{y_e=0} = 6.2\pm0.7({\rm stat.})\pm1.5({\rm syst.}) {\rm nb}.$$

FONLL can describe these measurements within its theoretical uncertainties. Future measurements on low-$p_T$ electrons from bottom meson decay are important to overcome the large uncertainties of the derived total bottom quark production cross section that originate mostly from the large variations of theoretical model prediction in the low-$p_T$ region. 

\begin {acknowledgments}
We thank the RHIC Operations Group and RCF at BNL, the NERSC Center at LBNL and the Open Science Grid consortium for providing resources and support. This work was supported in part by the Offices of NP and HEP within the U.S. DOE Office of Science, the U.S. NSF, the Sloan Foundation, the DFG cluster of excellence `Origin and Structure of the Universe'of Germany, CNRS/IN2P3, FAPESP CNPq of Brazil, Ministry of Ed. and Sci. of the Russian Federation, NNSFC, CAS, MoST, and MoE of China, GA and MSMT of the Czech Republic, FOM and NWO of the Netherlands, DAE, DST, and CSIR of India, Polish Ministry of Sci. and Higher Ed., Korea Research Foundation, Ministry of Sci., Ed. and Sports of the Rep. Of Croatia, and RosAtom of Russia.
\end {acknowledgments}


\end{document}